\documentstyle[pre,multicol,aps,amssymb,amstex]{revtex}

\begin{document}
\setlength{\textwidth}{180mm}
\setlength{\textheight}{240mm}
\setlength{\parskip}{2mm}
\input{epsf.tex}
\epsfverbosetrue
\renewcommand{\baselinestretch}{2.0}
\draft
%\epsfverbosetrue
\title{Optical Vortex Solitons in Parametric Wave Mixing}

\author{Tristram J. Alexander$^{1}$, Yuri S. Kivshar$^{1}$, 
Alexander V. Buryak$^{1,2}$ and Rowland A. Sammut$^{2}$}

\address{$^{1}$Optical Sciences Center, Research School of Physical
Sciences and Engineering, Australian National University, \\
Canberra ACT 0200, Australia}

\address{$^{2}$School of Mathematics and Statistics, Australian Defence
Force Academy, Canberra ACT 2600, Australia}

\maketitle

\begin{abstract}
We analyze {\em two-component spatial optical vortex solitons} supported by 
parametric wave mixing processes in a nonlinear bulk medium. We study two 
distinct cases of such localised waves, namely, parametric  vortex solitons
due to phase-matched second-harmonic generation in an optical medium with 
competing {\em quadratic} and {\em cubic} nonlinear response, and vortex 
solitons in the presence of {\em third-harmonic generation} in a cubic
medium.  We find, analytically and numerically, the structure of
two-component vortex solitons, and also investigate modulational 
instability of their plane-wave background. In particular, we predict 
and analyze in detail {\em novel types} of vortex solitons, {\em a 
`halo-vortex'}, consisting of a two-component  vortex core surrounded by a 
bright ring of its harmonic field, and {\em a `ring-vortex'} soliton 
which is a vortex in a harmonic field that guides a ring-like localized 
mode of the fundamental-frequency field.
\end{abstract}

\section{Introduction}

An optical vortex soliton appears as a stationary self-trapped beam in a 
self-defocusing optical medium that carries a phase singularity on an 
electromagnetic field, so that the  beam intensity vanishes at a certain 
point, and the field phase changes by  $2\pi m$ ($m$ being integer) along 
any closed loop around the zero-intensity
point.  If such an object is created in a linear bulk medium 
\cite{nye,nye_book}, it 
preserves the singularity but  expands due to diffraction.   
However,
in a nonlinear medium, the diffraction-induced  expansion of the vortex 
core  can be  compensated for by a nonlinearity-induced change in the 
refractive index of a nonlinear medium, thereby creating a stationary 
self-trapped  structure,  {\em an optical vortex soliton.}  Such nonlinear 
localized waves carrying a singularity
were first introduced as stationary solutions of the nonlinear
Schr{\"o}dinger (NLS) equation in the pioneering paper by Ginzburg and
Pitaevsky \cite{gin_pit} to describe topological excitations in
superfluids, but the same objects appear in many other fields \cite{pismen} 
including nonlinear optics \cite{dark_review}.

The parametric interactions may provide an efficient way of vortex 
transformation.  In particular, by mixing
waves of different frequencies, one can  change the vortex  topological
charge $m$ and even the vortex polarization.  Recently, the first experimental
results on the vortex generation in the presence of two-wave parametric
mixing have been reported in nonlinear optics, including the second-harmonic 
generation (SHG) \cite{soskin,dho} and more general types of frequency 
conversion \cite{berz} and sum-frequency mixing \cite{lithuania} where the 
generation
of higher-order ($|m| >1)$ linear vortices in the case of negligible
spatial walk-off between harmonics was demonstrated.

To the best of our knowledge, no theory of parametric optical vortices 
in
the presence of both diffraction and nonlinearity has been developed so
far. In a nonlinear regime, an interplay between diffraction and
parametric coupling of the harmonic fields is expected to lead to the 
formation of stationary structures - 
{\em parametric vortex solitons} - supported by three- or four-wave 
mixing between
the phase-matched waves of different frequencies.  Stability of such multi-
frequency
vortex
solitons is a key issue. For example, in the problem of
SHG in a diffractive bulk medium,  vortex
solitons
are expected to be unstable due to parametric modulational instability
of
the two-wave background field \cite{MI_bk}. Recently, it has been
suggested
\cite{tristram} that taking into account {\em a weak defocusing cubic
nonlinearity} one can eliminate the
development of parametric modulational instability allowing stable dark
solitons
to exist. Some examples of stable two-wave parametric dark
solitons have been presented in Ref. \cite{tristram}, and it has
been pointed out that, in the problem of SHG, a stable vortex soliton of 
the lowest possible charge
($|m| = 1$) can exist describing a $2\pi$-phase twist of
the fundamental wave and $4\pi$-phase twist in the second-harmonic field.

In the present paper we suggest a general approach to the analysis of 
{\em multi-component vortex solitons} resulting from parametric wave mixing.
The general theory is then developed in detail in the no-walkoff case for 
{\em two examples}:
(i) parametric interaction of the first and second
harmonics in a medium with competing quadratic and cubic nonlinearity, 
and
(ii) parametric interaction between the first and third
harmonics in a medium with a cubic nonlinear response.
In both the cases we find different classes of vortex solitons as (2+1)-
dimensional dark solitons
of circular symmetry carrying a phase singularity,
and investigate their stability to propagation and modulational stability of 
the supporting two-component background waves.

The paper is organized as follows. In Sec. II we briefly present two models of
parametric wave interaction that describe a phase-matched coupling
between
the fundamental frequency mode and its harmonic field,  in the case of 
phase-matched wave mixing and no walk-off. The further analysis of the
asymptotic structure of stationary localized solutions for parametric
vortex solitons is rather general,
and it is presented in Sec. III for both the models. Section IV is
devoted
to the analysis of vortex solitons in the model of competing
nonlinearities.
We find numerically the profiles of two-component vortex solitons and
investigate their stability to propagation. In particular,
we
reveal the existence of  novel classes of dark-soliton solutions of
radial
symmetry, including  {\em a ring-vortex soliton}, that consists of a
vortex
core in the harmonic field surrounded by a bright ring of its
fundamental
frequency, and
{\em a halo-vortex}, a two-wave vortex soliton with nonmonotonic tails.
The corresponding results are also obtained for the problem of
the third-harmonic generation in Sec. V.  Finally, Sec. VI gives the 
summary of our results and briefly discusses some related issues including 
the comments on experimental verifications and a link with other problems.

\section{Models of two-wave parametric interaction}

\subsection{Competing Nonlinearities}

First, we consider the model of competing quadratic and cubic
nonlinearities introduced earlier for the (1+1)-dimensional case
in Ref. \cite{compet} and recently generalized to the case of 
(2+1)-dimensional bright
solitons of radial symmetry in a bulk medium \cite{compet_3D}. 
We assume that a beam of a
fundamental harmonic (FH) with the frequency $\omega$ is launched into a 
medium possessing combined
quadratic [or $\chi^{(2)}$] and cubic [or $\chi^{(3)}$] nonlinear
response under the condition of phase-matched SHG.
The FH beam generates a second harmonic (SH) wave, and such a
two-wave mixing process in a bulk medium is described
by a system of two coupled nonlinear equations,
\begin{equation}
\label{compphys}
\begin{array}{l}
{\displaystyle 2ik_{1}\frac{\partial E_{1}}{\partial z} +
\nabla^{2}_{\perp} E_{1} + \frac{8\pi\omega^{2}}{c^{2}}\chi^{(2)}
E_{2}E_{1}^{*}e^{-i\Delta kz}
+ } \\*[9pt]
{\displaystyle
\frac{12\pi\omega^{2}}{c^{2}}\chi^{(3)}\left(\left| E_{1}\right|^{2}+
\rho\left|E_{2}
\right|^{2}\right) E_{1} = 0,} \\*[9pt]
{\displaystyle 4ik_{1}\frac{\partial E_{2}}{\partial z} +
\nabla^{2}_{\perp}
E_{2}
+ \frac{16\pi\omega^{2}}{c^{2}}\chi^{(2)}E_{1}^{2}e^{i\Delta kz}
+} \\*[9pt]
{\displaystyle
\frac{48\pi\omega^{2}}{c^{2}}\chi^{(3)}\left(\left| E_{2}\right|^{2}+
\rho\left|E_{1}
\right|^{2}\right) E_{2} = 0,} \\*[9pt]
\end{array}
\end{equation}
where $E_{1}$ and $E_{2}$ are the complex amplitude envelopes of FH
($\omega_{1} = \omega$) and SH ($\omega_{2} = 2\omega$) waves, 
respectively; $k_{1} = k(\omega)$ and $k_{2} = k(2 \omega)$ are the 
corresponding wave numbers; $\Delta k \equiv (2k_{1} - k_{2})$ is the 
wave-vector 
mismatch between the harmonics, $\rho$ (which we take $\rho = 2$) is 
the cross-phase-modulation coefficient, and the coefficients $\chi^{(2)}$ and 
$\chi^{(3)}$ are proportional to the second- and third-order susceptibility 
tensor elements and they
characterize the combined nonlinear response of an optical medium.

Adopting a similar set of scaling transformations as in Ref.
\cite{compet_3D}, we measure the transverse coordinates in the units of 
the
beam radius $R_{0}$, and the propagation coordinate, in the units of the 
beam diffraction length $R_{d} = 2k_{1}R_{0}^{2}$.  Then, applying the 
transformations
\begin{equation}
\label{scaling}
\nonumber
\begin{array}{l}
{\displaystyle E_{1} = \beta
c^{2}(16\pi\omega^{2}\chi^{(2)}R^{2}_{0})^{-1} e^{i\beta z} u(x,y,z),}
\\*[9pt]
{\displaystyle E_{2} = \beta
c^{2}(8\pi\omega^{2}\chi^{(2)}R_{0}^{2})^{-1} e^{i[(2\beta +
\Delta)z]} w(x,y,z),}
\end{array}
\end{equation}
where the parameter
$\beta$ stands for the nonlinearity-induced change of the beam
propagation
constant and $\Delta = 2 k_{1} R_{d}^{2} \Delta k$, we obtain a system 
of
normalized equations for $u$ and $w$,
\begin{equation}
\label{normalcomp}
\begin{array}{l}
{\displaystyle i\frac{\partial u}{\partial z} + s \nabla^{2}_{\perp} u- 
u +
wu^{\ast} + \chi \Big(\frac{|u|^{2}}{2\sigma} + \rho |w|^{2}\Big) u =
0,} \\*[9pt]
{\displaystyle i\sigma\frac{\partial w}{\partial z} + s 
\nabla^{2}_{\perp} w- \alpha w + \frac{u^{2}}{2} + \chi (2\sigma |w|^{2} 
+ \rho |u|^2) w = 0,}
\end{array}
\end{equation}
where $\alpha = (2\beta +\Delta)\sigma/\beta$, $s \equiv {\rm
sign} \beta$, and the coordinates are rescaled as follows $z \rightarrow
z/\beta$ and $(x,y) \rightarrow (x,y)/\sqrt{|\beta|}$.  For the
spatial beam propagation we take $\sigma = 2$.  Parameter
$\chi$ describes a competition between quadratic and cubic
nonlinearities,  and it is defined as
\begin{equation}
\label{chi}
\chi = \beta \frac{3
c^{2}}{16\pi\omega_{1}^{2}R_{0}^{2}}\frac{\chi^{(3)}}
{[\chi^{(2)}]^{2}}.
\end{equation}

Stationary solutions are then described by Eqs. (\ref{normalcomp})
with the $z$-derivatives omitted.
To look for radially symmetric solutions carrying
a phase singularity, we use
the polar coordinates $r = \sqrt{x^{2}+y^{2}}$, $\phi = 
\tan^{-1}(x/y)$, and make the following substitutions,
\begin{equation}
\label{ansatz}
u(r,\phi) = U(r) e^{im \phi}, \;\;\;  w(r,\phi) = W(r) e^{i N m 
\phi},
\end{equation}
where $U(r)$ and $W(r)$ are real functions and, for parametric interaction 
between the fundamental
and second harmonics, $N = 2$ whereas $m$ is an integer number that 
characterises the vortex charge.

Substituting Eqs. (\ref{ansatz}) into Eqs. (\ref{normalcomp}), we
obtain
\begin{equation}
\label{station}
\begin{array}{l}
{\displaystyle \frac{d^{2} U}{d r^{2}} + \frac{1}{r}\frac{d U}{d r} -
\frac{m^{2} U^{2}}{r^{2}} + s \frac{\partial F}{\partial U} =0,}
\\*[9pt]
{\displaystyle \frac{d^{2} W}{d r^{2}} + \frac{1}{r}\frac{d W}{d r} -
\frac{m^{2} N^{2} W^{2}}{r^{2}} + s \frac{\partial F}{\partial W} =0,}
\end{array}
\end{equation}
where the function $F$ has the meaning of an effective potential, and 
it is defined as
\begin{equation}
\label{F1}
\begin{array}{l}
{\displaystyle F = F_{1}(U,W) = - \frac{1}{2} U^{2} +
\frac{1}{2} U^{2} W - \frac{\alpha}{2}W^{2}  } \\*[9pt]
{\displaystyle  + \chi \Big(\frac{1}{16} U^4 + W^4  +
\frac{1}{2} \rho W^2 U^2 \Big).}
\end{array}
\end{equation}

\subsection{Third-Harmonic Generation}

A similar type of two-wave parametric interaction occurs
under the condition of the third-harmonic generation (THG).
Bright and dark solitary waves in a waveguide geometry
(i.e. with one transverse dimension) have been analyzed
in Ref. \cite{thg}.  In this case, the parametric interaction occurs 
between the fundamental beam ($\omega_{1} = \omega$)
and its third harmonic ($\omega_{3} = 3 \omega$), and the corresponding
physical model of the parametric wave mixing in a bulk can be described by
a system of two 
coupled equations,
\begin{equation}
\label{thgphys}
\begin{array}{l}
{\displaystyle 2 ik_{1}\frac{\partial E_{1}}{\partial z} + s \nabla^{2} E_{1} -
} \\*[9pt]
{\displaystyle \chi \left[\left(|E_{1}|^{2} + 2 |E_{3}|^{2}\right) E_{1} +
E_{1}^{\ast 2}E_{3} e^{-i\Delta kz}\right] = 0,} \\*[9pt]
{\displaystyle 2ik_{3}\frac{\partial E_{3}}{\partial z} + s \nabla^{2} E_{3} -
} \\*[9pt]
{\displaystyle 9\chi \left[\left(|E_{3}|^{2} + 2 |E_{1}|^{2}\right) E_{3} +
\frac{1}{3}E_{1}^{3}E_{3} e^{i\Delta kz}\right] = 0,} \\*[9pt]
\end{array}
\end{equation}
where $E_{1}$ and $E_{3}$ are the slowly varying envelopes of the 
first and third harmonic fields, respectively, with corresponding 
wave numbers $k_{1} = k(\omega)$ and $k_{3} = k(3\omega)$; 
$\Delta k = 3k_{1}-k_{3}$ is the wave-vector mismatch between the harmonics 
and $\chi = (3\pi \omega^{2}/c^{2})|\chi^{(3)}|$ is the nonlinearity 
parameter, which is assumed here to be always positive, whereas 
$\chi^{(3)} < 0$.

We follow a normalisation procedure similar to that used above for the 
competing nonlinearity model.  Again, the transverse coordinate is measured 
in units of the beam width $R_{0}$ and the propagation coordinate, in units of 
the diffraction length $R_{d} = 2k_{1}R_{0}^{2}$.  Using the transformations 
of 
Ref. \cite{thg}
\begin{equation}
\label{thgtrans}
\begin{array}{l}
{\displaystyle E_{1} = \left(\sqrt{\beta}/3\sqrt{k_{1}R_{0}^{2}\chi}\right)
e^{i\beta z}
u(x,y,z),} \\*[9pt]
{\displaystyle E_{2} = \left(\sqrt{\beta}/\sqrt{k_{1}R_{0}^{2}\chi}\right)
e^{i(3\beta +\Delta)z}w(x,y,z),} \\*[9pt]
\end{array}
\end{equation}
the physical equations (\ref{thgphys}) can be written in the following 
normalised form [cf. Eqs. (\ref{normalcomp})],
\begin{equation}
\label{thgnormal}
\begin{array}{l}
{\displaystyle i\frac{\partial u}{\partial z} + s \nabla^{2} u - u -
\frac{s}{3} u^{\ast 2}w - s \left(\frac{|u|^{2}}{9} + 2 |w|^{2}\right) u 
= 0,}
\\*[9pt]
{\displaystyle i\sigma\frac{\partial w}{\partial z} + s \nabla^{2} w
- \alpha w - \frac{s}{9}u^{3} - s (9 |w|^{2} + 2 |u|^2) w = 0,}
\end{array}
\end{equation}
where $u$ and $w$ are the normalised amplitudes of the fundamental harmonic
field and its third harmonic,
$\alpha = \sigma (3 \beta + \Delta)/\beta$, $\Delta =
2k_{1}R_{0}^{2}\Delta k$,
$s \equiv {\rm sign} \beta$, the transverse and propagation coordinates have 
been rescaled 
in terms of the nonlinearity-induced change of the propagation constant $\beta$, 
$z \rightarrow z/\beta$ and $(x,y) \rightarrow (x,y)/\sqrt{|\beta|}$, and, for 
spatial solitons, we take $\sigma = 3$.
Importantly, everywhere below we consider only
{\em defocusing cubic nonlinearity} searching for vortex-type
solitary waves on a modulationally stable nonvanishing background.

Stationary radially symmetric localized solutions of
Eqs. (\ref{thgnormal}) have the form (\ref{ansatz}) with $N = 3$, and 
they
satisfy Eqs. (\ref{station}) with the potential $F$, this time  
defined as
\begin{equation}
\label{F2}
\begin{array}{l}
{\displaystyle F = F_{2}(U,W) = - \frac{1}{2} U^{2} - 
\frac{\alpha}{2} W^{2} } \\*[9pt]
{\displaystyle  - s \Big(\frac{1}{9} U^{3} W +
 \frac{1}{36} U^4 + \frac{9}{4} W^4  +  W^2 U^2 \Big).}
\end{array}
\end{equation}

Thus, in both the cases, stationary vortex-like structures
are described by the same system of equations (\ref{station})
with different types of the potential $F$.
This observation allows us to perform further analytical
calculations in a rather general form, and, therefore, most of them 
are universal and can be applied to other models.

\section{General theory of parametric vortex solitons}

\subsection{Stationary Solutions}

Stationary radially symmetric solutions of Eqs. (\ref{normalcomp}) [Eqs. 
(\ref{thgnormal})] are given by Eqs. (\ref{station}) with the potential 
function  $F$ defined in Eq. (\ref{F1}) [Eq. (\ref{F2})] and $N = 2$ 
[$N = 3$]. It is important to note that the parametric coupling 
between the modes brings {\em several new features} in the vortex 
structure and properties. Indeed, as follows from Eqs. (\ref{ansatz}) 
and  (\ref{station}), a vortex with the charge $m$ in the fundamental
mode is always coupled to a vortex of the charge $N m$ ($N = 2,3$) in 
the harmonic component. This makes parametric vortices very different 
from
all types of vortex solitons analyzed earlier in the systems of two 
incoherently coupled NLS equations (see, e.g., Ref. \cite{sheppard1} and 
references therein).

\subsection{Analysis of Vortex Asymptotics}

We are interested in the localized solutions supported by a two-component
finite-amplitude background wave.  For $r \rightarrow \infty$, the background 
amplitudes ($U_0,W_0$) satisfy the coupled algebraic equations:
\begin{equation}
\label{back}
\frac{\partial F}{\partial U} = 0,  \hspace{1cm} \frac{\partial
F}{\partial W} = 0,
\end{equation}
which may have one or more nontrivial solutions.
Importantly, due to the self-action effect we always have a special 
solution of the form
$(0,W_0)$, that corresponds to an excited harmonic field only.

A vortex soliton is a localized nonlinear
mode that asymptotically approaches the background
$(U_0,W_0)$ for $r \rightarrow \infty$, but its intensity
vanishes for $r \rightarrow 0$ to keep the terms
$\sim (m^{2}/r^{2}) U$ and $\sim (m^{2} N^{2} /r^{2}) W$
in Eqs. (\ref{station}) finite.
This implies that we can find the vortex asymptotics in a rather 
general form. For $r \rightarrow 0$, we look for solutions
of Eqs. (\ref{station}) in the form:
\begin{equation}
\label{AABB}
\begin{array}{l}
{\displaystyle U = U_{0} - \frac{A}{r^{2}} - \frac{A_{2}}{r^{4}} +
\ldots,} \\*[9pt]
{\displaystyle W = W_{0} - \frac{B}{r^{2}} - \frac{B_{2}}{r^{4}} +
\ldots,} \\*[9pt]
\end{array}
\end{equation}
where $(U_0,W_0)$ is a solution of Eqs. (\ref{back}) for the background
amplitudes.
Keeping in Eqs. (\ref{station}) only the asymptotic terms up to the order 
of $\sim 1/r^{2}$, 
we obtain,
\begin{equation}
\label{a_b}
\begin{array}{l}
{\displaystyle  s m^{2} U_{0}
+ \Big(\frac{\partial^{2} F}{\partial U^{2}}\Big)_{0} A
+ \Big(\frac{\partial^{2} F}{\partial U \partial W}\Big)_{0} B =0,}
\\*[9pt]
{\displaystyle s N^{2} m^{2} W_{0}
+ \Big(\frac{\partial^{2} F}{\partial W^{2}}\Big)_{0} B
+ \Big(\frac{\partial^{2} F}{\partial W \partial U}\Big)_{0} A =0,}
\end{array}
\end{equation}
where the index $'0'$ stands for the values calculated at $U = U_0$
and $W = W_0$. Solutions of the linear equations (\ref{a_b}) for $A$ 
and $B$ can  be easily found analytically; they define the 
asymptotics of the vortex solitons for different values of the vortex 
charge $m$ in terms of the background amplitudes $U_{0}$ and $W_{0}$ 
defined by Eqs. (\ref{back}).

The analysis of the asymptotics gives us important information
about the vortex structure. If both the products $A U_0$ and $B W_0$  are 
positive [see Eqs.  (\ref{AABB})], the vortex has a standard profile with 
the intensity in the core growing monotonically and always lower 
than the background intensity. However, if one of these products
is negative, somewhere across the vortex the intensity becomes higher 
than the asymptotic background intensity. That implies that the vortex 
core is surrounded by a bright ring of higher intensity. We call 
such structures {\em `halo-vortices'}. In both the cases  mentioned 
above, such  vortex solitons may exist on a modulationally stable 
background, and some  examples are given below in Sections \ref{comp_non} 
and \ref{third_non}.

\subsection{Vortex Soliton as a Waveguide}

The concept of light guiding light (see e.g., Ref. \cite{phys_today} and
references therein) is based on a simple observation that a spatial optical 
soliton 
(e.g., vortex)
creates an effective optical waveguide in a nonlinear medium that can
guide a wave of different frequency or polarization. It is
clear that a vortex soliton creates a waveguide of radial symmetry
which  can guide a fundamental mode (no nodes) of the other wave. For 
the case of two incoherently coupled NLS equations describing two
orthogonal polarizations, the guiding properties of vortex solitons
have been analyzed by Haelterman and Sheppard
\cite{sheppard1}.  The first demonstration of an optically written waveguide 
based on an optical vortex has been recently reported by Truscott {\em et al}
\cite{experiment}.  However, the theory developed in Ref. 
\cite{sheppard1}
is not valid for the case of the resonant interactions and parametrically 
coupled waves. 
Indeed, the parametric interaction forces
the harmonic field to vanish for $r \rightarrow 0$, {\em trapping a singularity} 
of the order of $N m$. Therefore, a parametric vortex
{\em cannot guide a fundamental mode}. To analyse the guiding properties 
of
parametric vortex solitons, we note that Eqs. (\ref{back})
with the potential $F$ defined by Eqs. (\ref{F1}) and (\ref{F2})
have the solution ($U_{0} = 0$, $W_{0} \neq 0$).
Therefore, we consider a vortex soliton created by
a harmonic field $W$, with a stationary profile described
by the nonlinear equation,
\[
\frac{d^2 W}{d r^{2}} + \frac{1}{r} \frac{d W}{d r} -
\left(\frac{N^2 m^2}{r^2}  + s \alpha\right)  W - \gamma  W^3 
= 0,
\]
where $\gamma = - 4 s \chi$, for the model (\ref{normalcomp}),
and $\gamma = 9$, for the model (\ref{thgnormal}).
This equation always has a solution in the form of a vortex
soliton with the charge $N m$ provided $\gamma > 0$ and $s \alpha < 0$.
Now,  an eigenvalue equation
for a linear mode guided by the vortex $W(r)$ follows from
the first equation of the system (\ref{station}). Assuming
$U \ll {\rm max}(W)$, we obtain,
\begin{equation}
\label{guide}
\frac{d^2 U}{d r^{2}} + \frac{1}{r} \frac{d U}{d r} -
\left[\frac{m^2}{r^2} +  s -  s G(r)\right] U = 0,
\end{equation}
where
\begin{equation}
\nonumber
G(r) = \Big(\frac{d^2 F}{d U^{2}}\Big)
\bigg|_{{U=0}, \;}.
\end{equation}

Equation (\ref{guide}) is a standard eigenvalue problem of
the linear waveguide theory,  and it can be studied analytically, e.g.
by means of variational methods (see, e.g.,
Ref. \cite{snyder_love} and references therein). To make some analytical 
estimates, we present $G(r)$ in  an approximate form and obtain
\[
\frac{d^2 U}{d r^{2}} + \frac{1}{r} \frac{d U}{d r} -
\frac{m^2}{r^2} U -  E U  +  \frac{C^2}{(D^2 + r^{2})} U = 0,
\]
where $E$, $C$, and $D$ are, in general, functions of $\alpha$ and 
$\gamma$.  The parameters are chosen to provide the best 
approximation
of the effective potential $G(r)$. Using the standard variational method (or 
Ritz optimisation approach) and looking for a bifurcation of a linear mode 
taken in a trial form, $f(r)  = r \exp{(-\kappa r)}$, we obtain an implicit 
expression to determine the mode cutoff
$\alpha$,
\begin{equation}
\nonumber
E = (2 C^2 - 1 - 2 m^2)^3/(36 C^4 D^2),
\end{equation}
which we analyse below for some particular cases.

\subsection{Modulational Instability}

Stability of the stationary vortex solitons described by the system 
(\ref{station}) is an important issue. In general, the stability analysis of 
vortices in nonlinear models is a complicated and, generally speaking, 
unsolved problem. Instability can develop due to the presence of unstable 
eigenmodes 
localized near the vortex core
and, in the one-dimensional case, this type of instability of
{\em dark solitons} leads to the soliton motion, i.e. it is {\em a drift
instability} (see, e.g., Ref. \cite{dark_review} and references 
therein).  Since moving vortices with nonzero minimum intensity (similar 
to grey solitons) do not exist, similar drift instability is not 
observed for vortices. The main instability which is usually associated 
with a vortex 
soliton originates from the instability of the nonlocalized background 
field.

The analysis of modulational instability of the background field can be 
carried out in a
general form. First, we write Eqs. (\ref{normalcomp}) and
Eqs. (\ref{thgnormal}) in the form
\begin{equation}
\label{nonstat}
\begin{array}{l}
{\displaystyle i \frac{d u}{d z} + s \nabla^{2} u +
\frac{\partial {\cal F}}{\partial u^{*}} =0,} \\*[9pt]
{\displaystyle i \sigma \frac{d w}{d z} + s \nabla^{2} w +
\frac{\partial {\cal F}}{\partial w^{*}} =0,}
\end{array}
\end{equation}
with ${\cal F}$ defined as
\begin{equation}
\label{complexF}
\begin{array}{l}
{\displaystyle {\cal F} \rightarrow {\cal F}_{1} =  - |u|^{2} +
\frac{1}{2} (u^{2} w^{*} + u^{*2} w) - \alpha |w|^{2} + } \\*[9pt]
{\displaystyle  \chi \Big(\frac{1}{8} |u|^4 + 2 |w|^4  +
\rho |w|^2 |u|^2 \Big),} \\*[9pt]
\end{array}
\end{equation}
for the model of competing nonlinearities, or
\begin{equation}
\begin{array}{l}
{\displaystyle {\cal F} \rightarrow {\cal F}_{2} =  - |u|^{2} +
\frac{s}{9} (u^{3} w^{*} + u^{*3} w) - \alpha |w|^{2} + } \\*[9pt]
{\displaystyle  + s \left(\frac{1}{18} |u|^4 + \frac{9}{2} |w|^4 
+  2 |w|^2 |u|^2\right),}
\end{array}
\end{equation}
for the model of the third-harmonic generation.
We look for stability
of the background wave solution $(U_{0},W_{0})$ defined by
Eqs. (\ref{back}), and linearize Eqs. (\ref{nonstat}) around
this stationary solution substituting:
\begin{equation}
\label{modulation}
\begin{array}{l}
{\displaystyle u = U_{0} + a e^{i {\bf \vec{k} \cdot  \vec{r}} + i \omega z} 
+
b e^{-i {\bf \vec{k} \cdot \vec{r}} - i \omega z},} \\*[9pt]
{\displaystyle w = W_{0} + c e^{i {\bf \vec{k} \cdot  \vec{r}} + i \omega z} 
+
d e^{-i {\bf \vec{k} \cdot \vec{r}} - i \omega z}.}
\end{array}
\end{equation}
As a result, we obtain a system of linear equations for $a$, $b^{*}$, 
$c$, and $d^{*}$ leading to the characteristic equation:
\begin{equation}
\label{modulat_equat}
\nonumber
\begin{vmatrix}
A_{u^{*},u^{*}}-\Omega & A_{u^{*},u} & A_{u^{*},w^{*}} & A_{u^{*},w} \\
A_{u,u^{*}} & A_{u,u} + \Omega & A_{u,w^{*}} & A_{u,w} \\
A_{w^{*},u^{*}} & A_{w^{*},u} & A_{w^{*},w^{*}}-\Omega & A_{w^{*},w} \\
A_{w,u^{*}} & A_{w,u} & A_{w,w^{*}} & A_{w,w} + \Omega \end{vmatrix}
= 0.
\end{equation}
Here $A_{n,m} \equiv (\partial^{2}{\cal F}/ \partial n \; \partial
m)|_{(u = U_{0}, w = W_{0})}$, where $m,n = (u, u^{*}, w, 
w^{*})$ ($n  \neq m$), and
for the $\alpha = \beta$ case
\[
A_{n,n} \equiv \left(\frac{\partial^2{\cal F}}{\partial n 
^2}\right)\big|_{(u = U_{0}, w  = W_{0})} - |\vec{\bf k}|^{2}.
\]
Solving the characteristic equation with respect to $\Omega$, we
conclude the modulational instability analysis: purely real $\Omega$ 
solutions for all positive $|\vec{\bf k}|^{2}$ (with other parameters fixed)
indicate a modulationally stable background for this fixed set of
the parameters.  Below we present the results of the modulational 
instability analysis for  two cases of the parametric two-wave 
interaction.

\section{Competing nonlinearities}
\label{comp_non}

\begin{figure}
\setlength{\epsfxsize}{12.0cm}
\centerline{\mbox{\epsffile{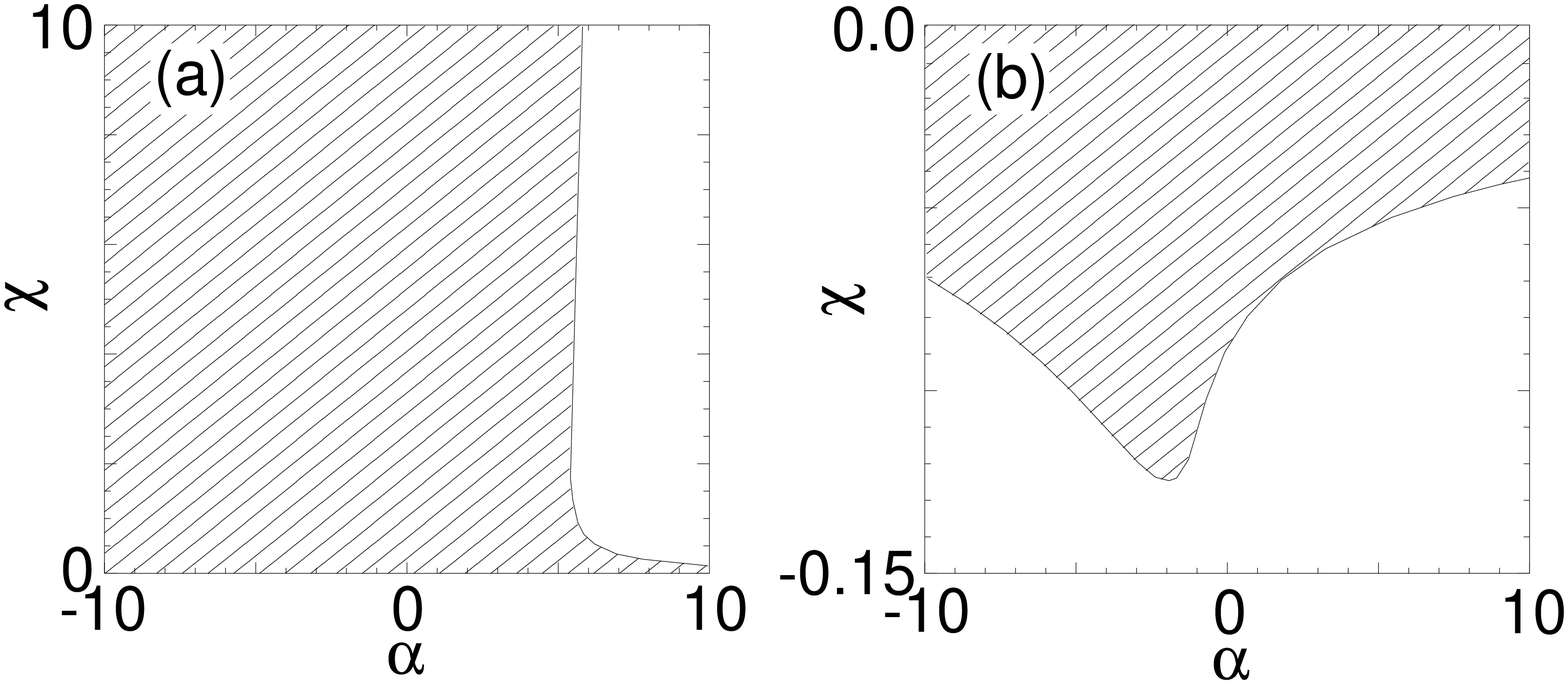}}}
\caption{Existence domains for the modulationally stable background
modes of the system (\protect{\ref{normalcomp}})
for (a)  $s = -1$ and $\chi > 0$; (b)  $s = +1$ and $\chi < 0$.}
\label{MI_comp}\end{figure}

Analysis of modulational instability for the system (\ref{normalcomp})
has been briefly presented in Ref. \cite{tristram}. Below we repeat the 
main steps of that analysis for the completeness of this paper.
Solutions of Eqs. (\ref{normalcomp}) for background waves can be found
by solving the coupled algebraic equations:
\[
\begin{array}{l}
{\displaystyle 12 \chi W_{0}^{3}+ 12  W_{0}^{2}+W_{0}(\alpha-8+ 2\chi^{-1}) =  
2\chi^{-1},} \\*[9pt]
{\displaystyle U_{0}^{2} = 4(1-W_{0})\chi^{-1} - 8 W_{0}^{2},}
\end{array}
\]
for real $U_{0}$ and $W_{0}$. There exist up to {\em three such solutions} 
with
both amplitudes $U_{0}$ and $W_{0}$ being nonzero. Performing  the 
analysis
of modulational instability for each of the three solutions at $s=\pm 
1$,  we find that there exists  only {\em one modulationally stable} 
mode. The~parameter domains where such a solution exists are 
presented in  Figs.~\ref{MI_comp}(a,b). Note, that $\rm sign (s \chi) = 
\rm sign \chi^{(3)}$ and thus modulationally stable solutions exist only 
for $\chi^{(3)} < 0$.
Importantly, the amplitude of the modulationally stable background 
diverge
in the limit $\chi \rightarrow 0$ so that the stable background solution
exists exclusively due to {\em mutual action of quadratic and cubic 
nonlinearities}. Other nonlinear modes are modulationally unstable in the 
whole domain of their existence and they are not presented in 
Figs.~\ref{MI_comp}(a,b).   Modulational stability in the limit of large 
negative $\chi^{(3)}$ [e.g., for $s= -1$ and $\chi >0$, see Fig. 
\ref{MI_comp}(a)], is not surprising because stable dark solitons are 
known to exist in a defocusing Kerr medium without quadratic 
nonlinearity.  Here,  we are  interested in the case when the effective 
nonlinearity is predominantly quadratic,  i.e.  $|\chi U_{0}| \sim |\chi 
W_{0}|~\ll~1$.  We  found that this condition can only be satisfied for 
$s = +1$ where modulationally stable background waves  of moderate 
amplitudes exist for relatively small values of negative $\chi$ [see Fig. 
1(b)].

\begin{figure}
\setlength{\epsfxsize}{12.0cm}
\centerline{\mbox{\epsffile{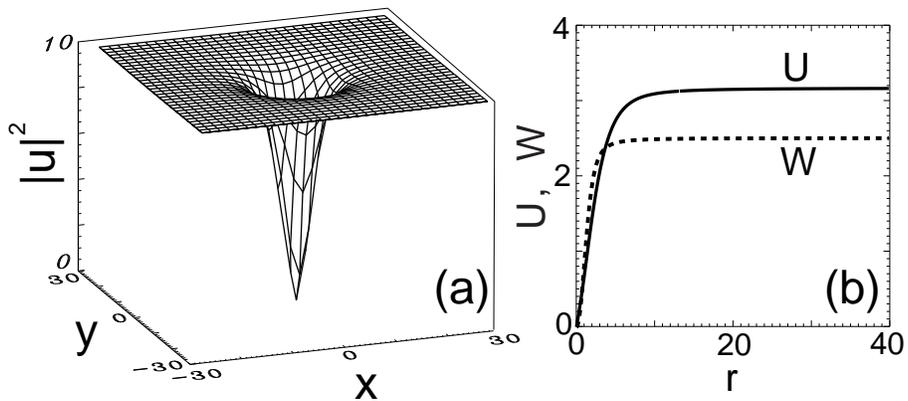}}}
\caption{An example of a two-component vortex soliton supported by competing 
nonlinearity  ($\alpha = -2.5$, $\chi  =-0.1$, and $s = +1$). In 
(a) only the vortex of the fundamental harmonic field is shown.}
\label{example}\end{figure}

Using the numerical relaxation technique, we have found that a 
continuous family of two-component vortex solitons exists in the whole region 
of the existence of modulationally stable background waves shown in 
Figs. \ref{MI_comp}(a, b). Figures \ref{example}(a) and \ref{example}(b) 
present an example of such a  vortex soliton.

\begin{figure}
\setlength{\epsfxsize}{12.0cm}
\centerline{\mbox{\epsffile{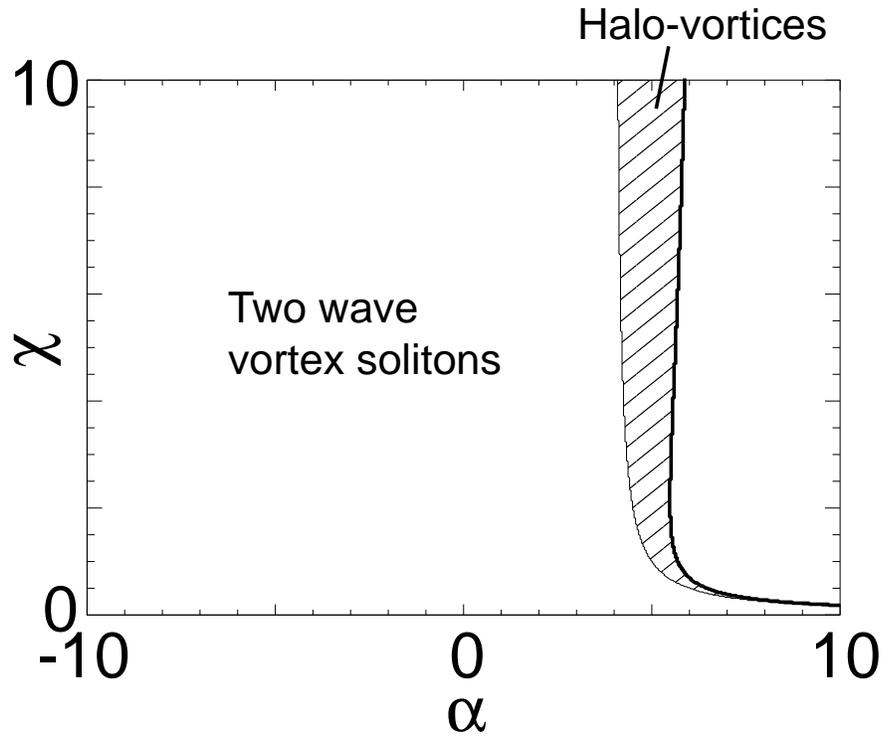}}}
\caption{Region of the existence of halo-vortices in the model of 
competing
nonlinearities, $s=-1$.}
\label{halocomp}\end{figure}

Analysis of the vortex asymptotics demonstrates that halo-vortices can 
exist  in Eqs. (\ref{normalcomp}) only if $s = -1$ and $\chi > 0$, in 
the narrow domain 
shown in Fig. \ref{halocomp}. Both terms with $B$ and
$B_{2}$ factors in the asymptotic expansion (\ref{AABB}) contribute to 
the formation of the halo (i.e. both $B W_0$ and $B_{2} W_0$ products 
are negative).

\begin{figure}
\setlength{\epsfxsize}{12.0cm}
\centerline{\mbox{\epsffile{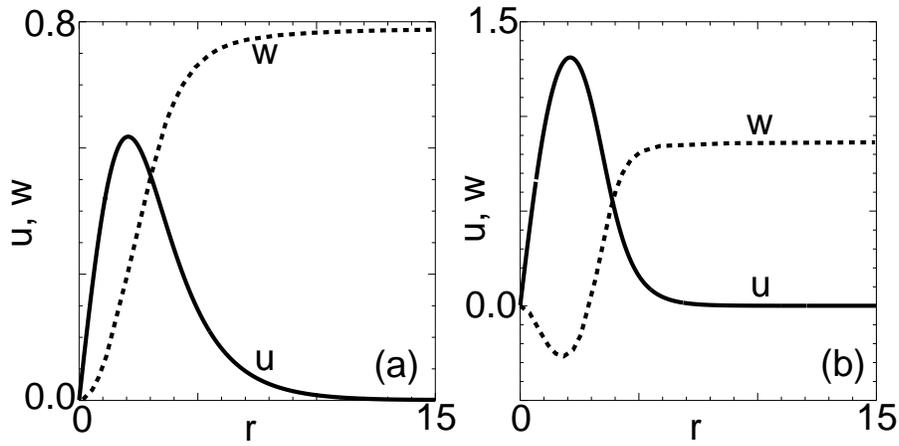}}}
\caption{Examples of a ring-vortex soliton ($s=-1$, $\chi = 1$) for
(a) $\alpha = 1.4$ and (b) $\alpha = 3$.  Solid - FH wave, dashed -
SH wave.}
\label{vb_examp_comp}\end{figure}

{\em Ring-vortex solitons} can also exist in  the model 
(\ref{normalcomp}), see Figs. 4(a,b) and Figs. 5(a,b). Variational 
analysis allows us to find
an approximate expression for the bifurcation curve where such solutions 
appear,
$(2-\alpha) =\sqrt{\alpha/\chi}$. As $\alpha$ increases,
the maximum of the bright-ring amplitude approaches the value $U_0$ of
two-wave modulationally stable parametric plane waves.
At the values of $r$ where $U$ approaches $U_0$, the second 
component,
$W$, also approaches the corresponding plane wave amplitude $W_0$ [see
Fig.  \ref{vb_examp_comp}(b)].

Such ring-vortex solitons can be unstable due to
modulational instability of the background wave $U_0 = 0$,
$W_0^2 = \alpha/4 \chi$.  For example, for $s = -1$, $\alpha >
0$, modulational stability is defined by the condition
$(\alpha - 2) > \sqrt{\alpha/\chi}$. The regions of the existence
of modulationally stable one-component plane waves and ring-vortex 
solitons of Eqs. (\ref{normalcomp})  are presented in Fig. \ref{major}.

Existence of two-component stable vortex solitons composed of 
parametrically coupled fields suggests that such vortices can be 
excited
in the process of the harmonic generation. In Fig. 7 we present the 
numerical simulation results supporting this idea.  We launch a mode of 
the fundamental frequency without a seeded second harmonic assuming the 
condition of phase-matching.  The vortex soliton dynamics is simulated 
using a split-step beam propagation method (BPM).  To solve the problem 
of the
vortex phase geometry, we simulate a system of four vortices, arranged 
such that horizontally and  vertically adjacent vortices are opposite in 
charge.  Lines of equal phase are chosen to correspond to the  lines of 
the force of an equivalent system of electrostatic point charges.  With 
the periodic  
boundary conditions imposed by BPM,  this configuration means that, in fact, 
an infinite array of vortices is simulated. Figure 7 shows the 
vortex generation by an input fundamental mode with single-charged 
vortices. Due to phase matching with the second harmonic, we observe a 
generation of double-charge vortices in the harmonic field (see the 
plots at $z=1$) and then periodic oscillations of the  two-component 
background and the vortex profiles near a stationary state corresponding 
to a lattice of two-component vortex solitons (see the plots at $z=10$ 
as an example of 
such dynamics).

\section{Third-harmonic generation}
\label{third_non}

Vortex solitons of Eqs. (\ref{thgnormal}) have fewer parameters in
comparison with the parametric vortices described by Eqs. (\ref{normalcomp}), 
and thus they can be analysed much more easily numerically.
These vortex solitons are found in the whole region
of the existence of modulationally stable plane waves,
i.e. for $\alpha < \alpha_{\rm th} \approx 14.509$.
Examples of such vortex solitons 
are shown in Figs. \ref{example_thg}(a, b).

A third-harmonic component of the vortex solitons
has a nonmonotonic tail for
$10.85 < \alpha < 14.509$, however it can only be called
a halo vortex for the interval $11.26 < \alpha < 14.509$, where
the absolute value of a local extremum in the structure of the
vortex tail is greater than the corresponding plane wave
background value $W_0$. The halo is becoming more
pronounced as $\alpha \rightarrow 14.509$.
An example of a halo-vortex soliton of Eqs. (\ref{thgnormal})
is shown in Fig. \ref{halo_thg}.

Ring-vortex solitons have also been found
for the model (\ref{thgnormal}), see Fig. 10.
In this case, a variational analysis allows us to find
an approximate analytical result for the bifurcation point ($s = -1$) 
where a ring-like mode is guided by the vortex:
\begin{equation}
\label{explicit}
\nonumber
\alpha_{\rm bif} = \frac{(\gamma/\delta)}{\{1-\frac{2}{9} \left[1-
\frac{\gamma}{2 \delta N^2 m^2} (1 + 2 m^2) \right]^3\}}.
\end{equation}
For $\gamma = 9$, $\delta = 2$, $N = 3$, and $m = 1$ this
gives $\alpha_{\rm bif} \approx 4.52$, which agrees well with
numerical data. We also find that, in general, coupled ring-vortex 
solitons exist for $\alpha > 4.5$, and for each such value of $\alpha$
(except $\alpha = \alpha_{\rm bif}$) there exist {\em two different types 
of
ring-vortex solitons} (see Fig. \ref{maximum}).  As the parameter $\alpha$ 
decreases,
the maximum of the bright ring in the fundamental mode approaches
the value of $U_0$ of the two-wave modulationally stable background field.
Again, as it has been observed for the model of competing 
nonlinearities, at values of $r$ where $U$ approaches $U_0$, the vortex 
component $W$ deforms significantly approaching the corresponding plane-wave
amplitude $W_0$. We note that {\em all these ring-vortex solitons are 
modulationally stable}, because, in the framework of Eqs. (\ref{thgnormal}), 
modulational instability does not occur for one-component plane wave solutions.

\section{Concluding Remarks}

We have analyzed two-component vortex solitons supported by parametric wave 
mixing in  a nonlinear optical medium. We have considered two classes of 
such vortex solitons.  In the first case, we have studied the existence, 
structure, and stability of vortex solitons supported by  
phase-matched interaction between  the
fundamental and second-harmonic waves in a quadratic medium, and the effect 
of the next-order cubic nonlinearity has been taken into account for 
suppressing modulational instability of the supporting plane-wave 
background. In the second case, we have considered how the vortex parameters, 
structure, and stability are modified due to the process of 
third-harmonic generation when the phase-matched wave interaction
generates a corresponding multi-charge vortex component in a harmonic 
field.  In particular,  we have predicted the so-called
`halo-vortex' consisting of a two-wave vortex core surrounded by a
bright
ring on a non-vanishing background.  Additionally, we have analyzed the
waveguiding properties of a vortex soliton in the case when it guides a
harmonic field due to a phase-matched parametric interaction.
A rigorous analysis of the stability of these parametric vortex solitons 
is
still an open problem, as well as the effect of walk-off on the vortex
existence and stability.

As for experimental verifications of the vortex solitons described above, 
we would like to mention that, at least in the low-intensity regime, 
parametric vortices have already been observed in nonlinear optics.  A 
possibility of SHG by a beam with a vortex was first mentioned and 
experimentally verified in Ref. \cite{soskin}, where a vortex of the 
topological charge $m=2$ was found in the second-harmonic wave when the 
fundamental wave contained a vortex of the topological charge $m=1$.  The 
similar results on SHG have been presented by Dholakia {\em et al} in 
Ref. \cite{dho}, whereas more complicated processes of sum-frequency 
mixing with beams carrying phase singularities were reported by 
Ber\v{z}anskis {\em et al} \cite{berz,lithuania}.  It is worth noticing that 
in all of 
those observations the different harmonics experienced noticeable walk-off 
that makes the stationary structures difficult to observe, also introducing 
novel features in the vortex dynamics.  In particular, for a collinear 
type I phase-matched SHG with an input beam carrying a single-charge vortex,
Matijo\v{s}ius {\em et al} \cite{matij} observed two intensity zeroes in 
the second-harmonic field with the separation of two SHG vortices due to 
walk-off.  Thus, we can expect that stationary two-component vortex solitons 
discussed above can be observed in typical upconversion experiments when a 
high-intensity beam undergoes frequency doubling simultaneously with the 
creation of a phase singularity produced by a phase mask at the input, 
similar to the experiments mentioned above which were performed at 
moderate powers.  {\em Stability} of those vortex solitons requires small 
(or zero) walk-off and a small defocusing Kerr nonlinearity of an optical 
material at both (or at least fundamental wave) frequencies.

Additionally, we would like to mention that the parametrically coupled 
equations 
of competing nonlinearities, similar to Eqs. (\ref{normalcomp}) analysed 
above, have been recently introduced by Heinzen {\em et al} \cite{heinzen} 
to describe the dynamics of {\em coupled atomic and molecular Bose-Einstein 
condensates}, leading to a kind of ``super-chemistry'' in which the 
formation of molecules is a controlled parametric quantum process.  In spite 
of the fact that both atomic and molecular condensates should be considered 
in a trapping external potential \cite{BEC}, many of the features of the 
coupled stationary states, including all the types of the vortex states 
introduced above, are expected to exist in the model of atom-molecular 
condensates as well, providing a much broader view of the phenomenology of 
parametric vortex solitons.

At last, we expect that the concept of the two-component parametric vortices, 
generated and supported by the third-harmonic generation process, can be 
important in the so-called {\em third-harmonic microscopy} (see, e.g., 
\cite{THGspectra}) where an image is rendered using a series of 
cross-sectional images produced by third-harmonic generation within the 
specimen.  Vortices can then be formed due to the development of caustics 
\cite{nye_book} in the reflected harmonic field, indicating the regions of 
highly concentrated inhomogeneities.  This technique is based on the fact that 
the nonlinear susceptibility of solid media vary over many orders of 
magnitude, compared with linear refractive index changes that vary in a 
relatively small range.

\section*{Acknowledgments}

Yuri Kivshar thanks O. Bang, P. DiTrapani, B. Ivanov, L. Pismen, A. 
Piskarskas, Y. Silberberg and V. Smilgevi\v{c}ius 
for useful discussions of the properties of multi-component and parametric 
vortices and harmonic generation.

The work has been supported by the Australian Photonics
Cooperative Research Centre and the Australian Research Council.  A 
brief
summary of the
results has been presented at the OSA Annual Meeting in
Baltimore, USA
(October 4-9, 1998).

\begin{figure}
\setlength{\epsfxsize}{12.0cm}
\centerline{\mbox{\epsffile{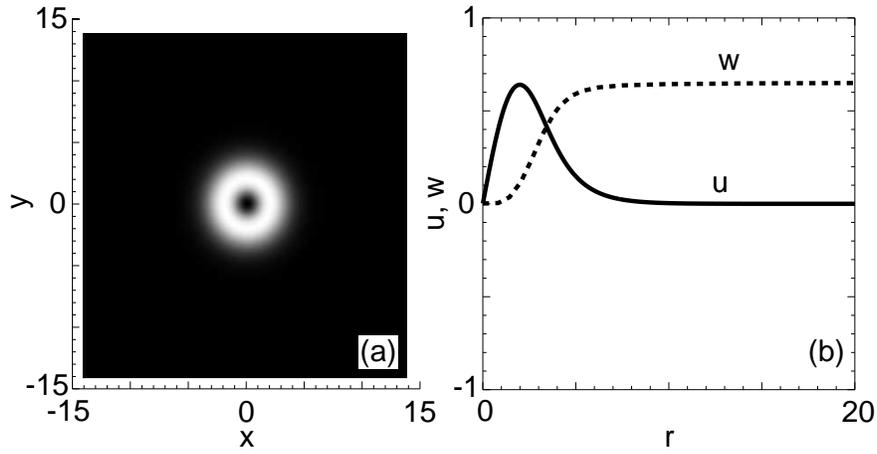}}}
\caption{(a) Intensity profile of the first harmonic, and (b) the
amplitude
profile of a ring-vortex soliton ($\alpha=1.7$, $\chi=1$, and 
$s=-1$) for the  fundamental (solid) and second-harmonic (dashed) 
modes.}
\label{vb_examp_comp2}
\end{figure}

\begin{figure}
\setlength{\epsfxsize}{12.0cm}
\centerline{\mbox{\epsffile{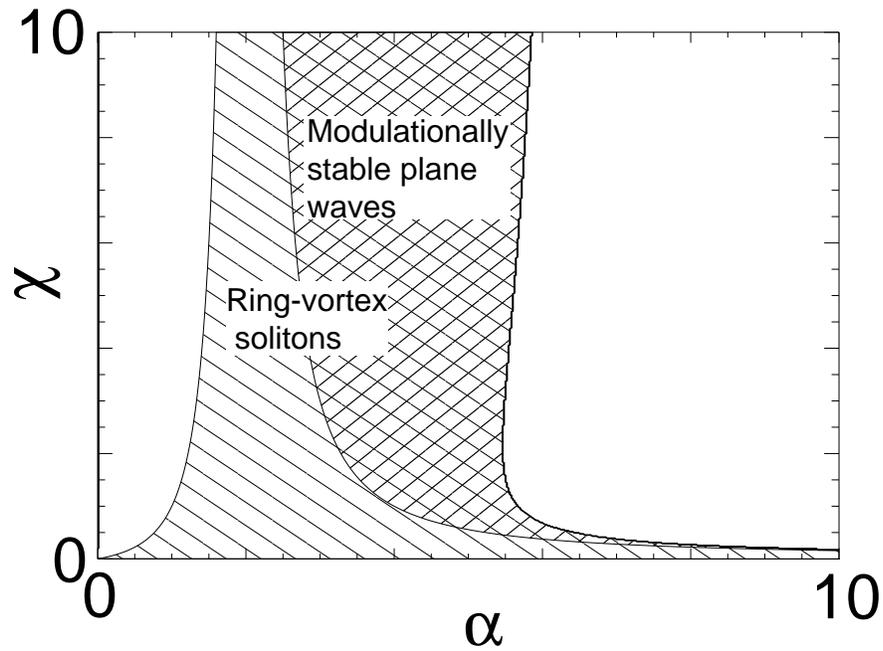}}}
\caption{Region of existence for ring-vortex solitons at $s = -1$ and 
$\chi > 0$.
Left curve, $(2-\alpha) =\sqrt{\alpha/\chi}$, is defined by 
a
bifurcation where a bright component appears.
Middle curve is a critical threshold that divides modulationally stable and 
unstable solutions. Right-hand side curve is the boundary for the existence of 
two-component parametric plane waves.}
\label{major}
\end{figure}

\begin{figure}
\setlength{\epsfxsize}{18.0cm}
\leftline{\mbox{\epsffile{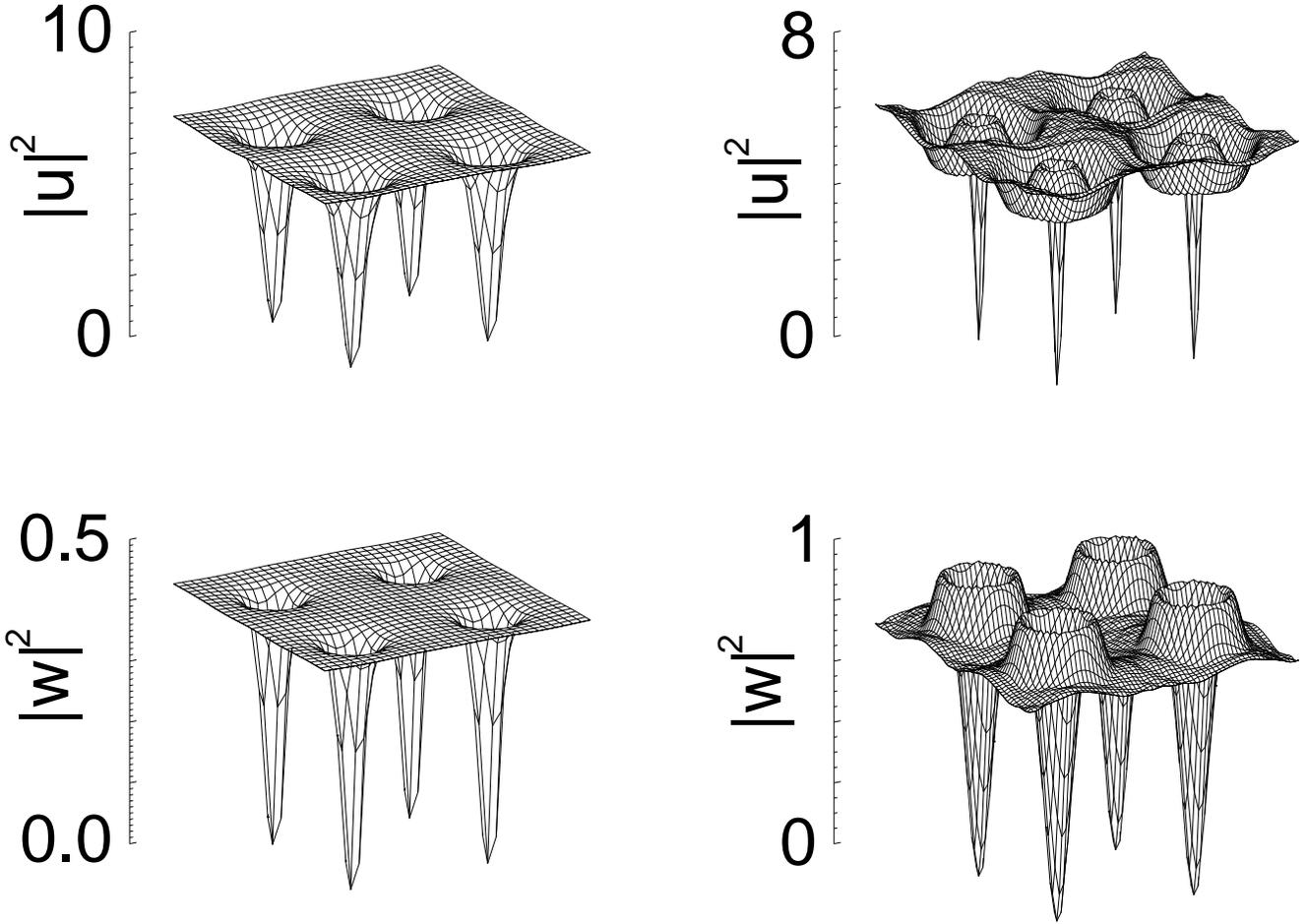}}}
\caption{Generation of two-component parametric vortex solitons
by the fundamental vortices. Profiles of the fundamental (upper row) and 
second-harmonic (lower row) fields are shown at $z=1$ (left column)
and $z=10$ (right column).}
\label{dynamics}
\end{figure}

\begin{figure}
\setlength{\epsfxsize}{12.0cm}
\centerline{\mbox{\epsffile{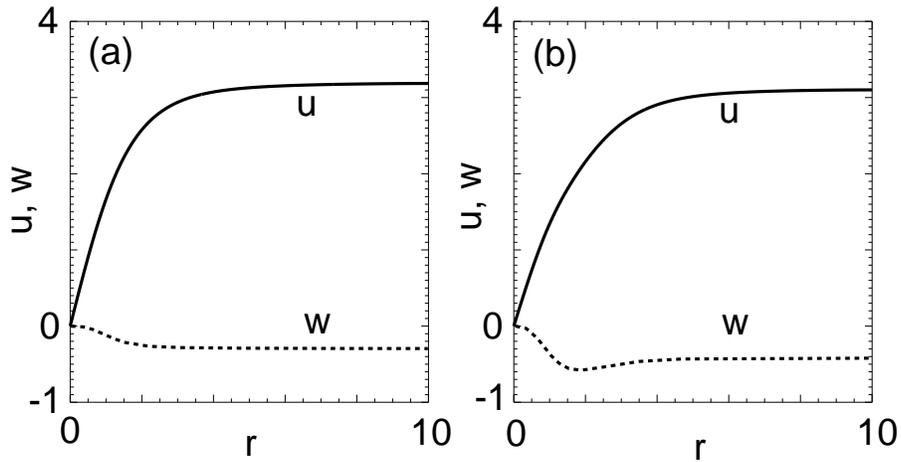}}}
\caption{Examples of two-component vortex solitons supported by the 
third-harmonic  generation at $s = -1$: (a) $\alpha = 9$, and (b)
$\alpha = 13$.}
\label{example_thg}\end{figure}

\begin{figure}
\setlength{\epsfxsize}{12.0cm}
\centerline{\mbox{\epsffile{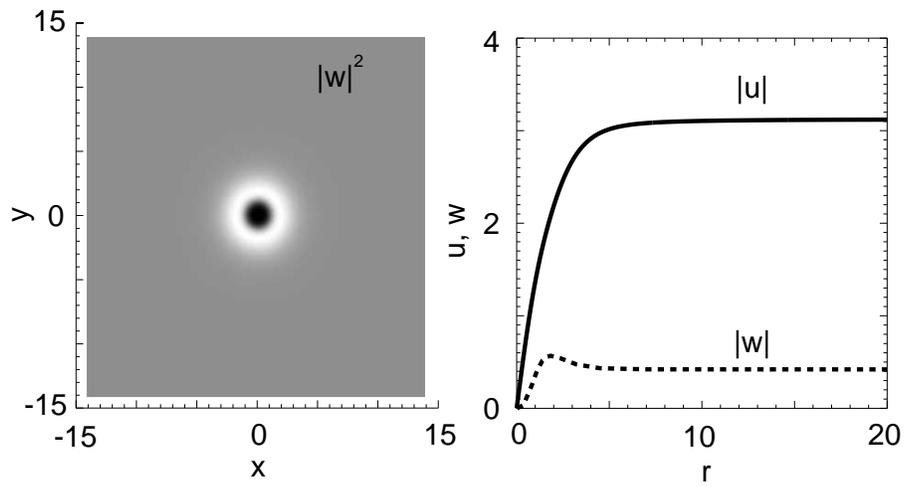}}}
\caption{Example of a halo-vortex soliton as a stationary solution of Eq. 
(\ref{thgnormal}) 
for $s = -1$ and $\alpha = 13$.}
\label{halo_thg}\end{figure}

\begin{figure}
\setlength{\epsfxsize}{10.0cm}
\centerline{\mbox{\epsffile{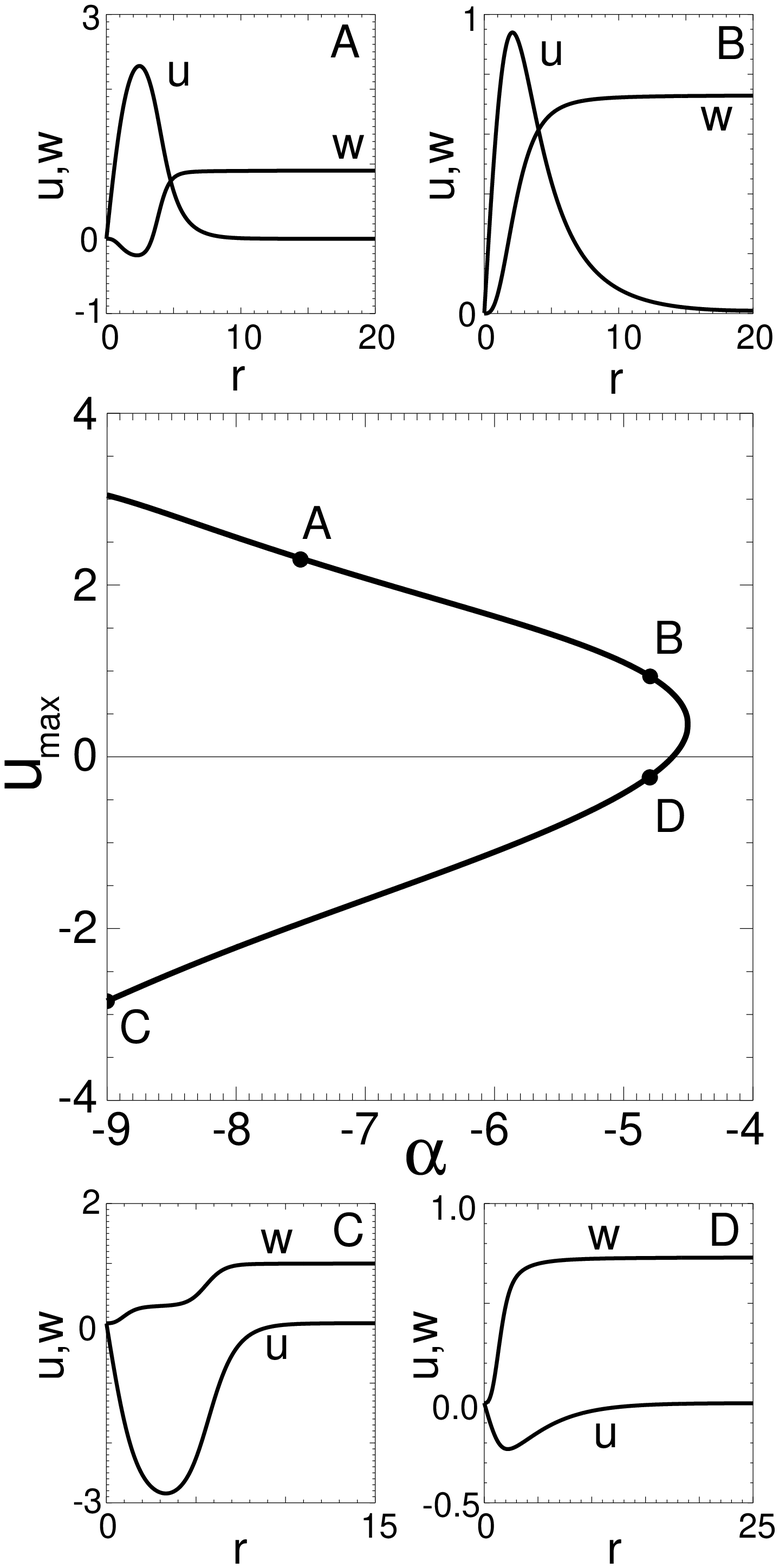}}}
\caption{A family of ring-vortex solitons shown in the middle plot 
as the maximum of
the bright-ring amplitude $u_{\rm max}$ vs. $\alpha$.  Filled circles A,
B, and C, D correspond to the vortex profiles shown above and below, 
respectively.}
\label{maximum}
\end{figure}


\begin{references}

\bibitem{nye} J.F. Nye and M.V. Berry, Proc. R. Soc. (London)
A {\bf 336}, 165 (1974).

\bibitem{nye_book} J.F. Nye, {\em Natural Focusing and Fine Structure of 
light:  Caustics and Wave Dislocations} (Institute of Physics Publishing, 
Bristol and Philadelphia, 1999)

\bibitem{gin_pit} V.L. Ginzburg and L.P. Pitaevsky, Zh. Eksp. Teor. Fiz.
{\bf 34}, 1240 (1958) [Sov. Phys. JETP {\bf 7}, 858 (1958)];  see also 
L.P.
Pitaevsky, Zh. Eksp. Teor. Fiz. {\bf 40}, 646 (1961)
[Sov. Phys. JETP {\bf 13}, 451 (1958)].

\bibitem{pismen} L.M. Pismen, {\em Vortices in Nonlinear Fields} (Oxford 
University Press, Oxford, 1999).

\bibitem{dark_review} See, e.g., Yu.S. Kivshar and B. Luther-Davies, 
Phys. Rep. {\bf 298}, 81 (1998).

\bibitem{soskin} I.V. Basistiy, V.Yu, Bazhenov, M.S. Soskin, and 
M.V. Vasnetsov, Opt. Comm. {\bf 103}, 422 (1993).

\bibitem{dho} K. Dholakia, N.B. Simpson, and M.J. Padgett, Phys. Rev. A
{\bf 54}, R3742 (1996).

\bibitem{berz} A. Ber\v{z}anskis, A. Matijo\v{s}ius, A. Piskarskas, V.
Smilgevi\v{c}ius, and A. Stabinis, Opt. Commun. {\bf 140}, 273 (1997).

\bibitem{lithuania} A. Ber\v{z}anskis, A. Matijo\v{s}ius, A. Piskarskas, 
V. Smilgevi\v{c}ius, and A. Stabinis, Opt. Comm. {\bf 150}, 372 (1998).

\bibitem{MI_bk} A.V. Buryak and Yu.S. Kivshar, Phys. Rev. A {\bf 51},
R41 (1995); A.V. Buryak and Yu.S. Kivshar, Opt. Lett. {\bf 20}, 834 
(1995); S. Trillo and P. Ferro, Opt.  Lett. {\bf 20}, 438 (1995).

\bibitem{tristram} T.J. Alexander, A.V. Buryak, and Yu.S. Kivshar,
Opt. Lett. {\bf 23}, 670 (1998).

\bibitem{compet} A.V. Buryak, Yu.S. Kivshar, and S. Trillo, Opt. Lett. 
{\bf 20}, 1961 (1995).

\bibitem{compet_3D} O. Bang, Yu.S. Kivshar,
A. V. Buryak, A. De Rossi, and S. Trillo, Phys. Rev. E
{\bf 58}, 5057 (1998).

\bibitem{thg} R.A. Sammut, A.V. Buryak, and Yu.S. Kivshar,
J. Opt. Soc. Am. B {\bf 15}, 1488 (1998).

\bibitem{sheppard1} See, e.g., M. Haelterman and A. P. Sheppard, Chaos,
Solitons and Fractals {\bf 4}, 1731 (1994); A. P. Sheppard and M. 
Haelterman, Opt. Lett. {\bf 19}, 859 (1994).

\bibitem{phys_today} M. Segev and G. Stegeman,
Physics Today {\bf 8}, 42 (1998).

\bibitem{experiment} A.G. Truscott, M.E.J. Friese, N.R. Heckenberg, and 
H. Rubinsztein-Dunlop, Phys. Rev. Lett. {\bf 82}, 1438 (1999).

\bibitem{snyder_love} A. W. Snyder and J. D. Love,
{\em Optical Waveguide Theory} (Chapman and Hall,
London, 1983).

\bibitem{matij} A. Matijo\v{s}ius, A. Piskarskas, V. Smilgevi\v{c}ius, 
and A. Stabinis, ``Parametric frequency conversion of optical vortices'', 
paper CWG5 of CLEO '99, OSA Technical Digest (OSA, Washington DC, 1999), 
p. 304.

\bibitem{heinzen} D.J. Heinzen {\em et al}, ``Super-chemistry: coherent 
dynamics of atom-molecular Bose condensates'', paper QTuD3 of QELS '99, 
OSA Technical Digest (OSA, Washington DC, 1999), p. 54.

\bibitem{BEC} See, e.g., F. Dalfovo, S. Giorgini, L.P. Pitaevskii, and 
S. Stringari, Rev. Mod. Phys. {\bf 71}, 
463 (1999).

\bibitem{THGspectra} Y. Barad, H. Eisenberg, M. Horowitz, and Y. Silberberg, 
Appl. Phys. Lett. {\bf 70}, 922 (1997).

\end{references}
\end{document}